\documentclass[runningheads]{llncs}
\usepackage[T1]{fontenc}
\usepackage{graphicx}
\usepackage{url}

\begin{document}

\title{A Role-Based Multi-Agent Model for Climate Adaptation Deliberation Across Living Labs}
\titlerunning{A Role-Based MACM for Climate Adaptation Deliberation}

\author{\"{O}nder G\"{u}rcan\inst{1} \and David Eric John Herbert\inst{2} \and F. LeRon Shultz\inst{1} \and \\ Christopher Frantz\inst{3} \and Ivan Puga-Gonzalez\inst{1}}
\authorrunning{G\"{u}rcan et al.}
\institute{NORCE Research AS, Kristiansand, Norway \and University of Bergen, Bergen, Norway \and NTNU, Gjøvik, Norway}

\maketitle

\begin{abstract}
Climate adaptation decisions involve heterogeneous stakeholders, partial information, and institutional constraints. This paper presents a role-based Multi-Agent Computer Model (MACM) to simulate climate adaptation deliberation across Living Labs. The main contribution is a configurable architecture that keeps behavioural mechanisms fixed while externalising context-specific inputs. Each stakeholder is represented as one agent that may play one or more roles: Expert Evaluator, Disseminator, Positioning Agent, and Decision-Maker. The simulation proceeds through four phases---initialization, information exchange, positioning and influence, and final decision-making. We argue that this role-based design improves both interpretability and cross-case reusability in social simulation of climate adaptation governance.
\end{abstract}

\section{Introduction}

Climate adaptation planning is shaped by uncertainty, competing priorities, and fragmented authority. Decisions are rarely made by a single actor following a simple optimisation rule. Instead, they emerge through deliberation among public authorities, domain experts, organised stakeholders, and other participants who differ not only in their preferences, but also in the function they perform within the process. Some actors produce assessments, some circulate information, some translate evidence into positions, and some hold formal decision authority. For social simulation, this implies that representing actors only as preference-bearing entities is often insufficient.

Existing models of participatory policy processes are frequently tailored to one empirical setting, with assumptions about stakeholder categories, information flows, and decision sequences embedded directly in the implementation. This limits transferability across cases and makes comparison difficult. Our starting point is that Living Labs (LLs) studying climate adaptation need a common simulation core that can travel across settings without erasing institutional differences\footnote{Here, LLs are participatory settings for examining climate adaptation proposals under local constraints.}. The challenge is therefore methodological: how can one design a model that remains reusable across cases while still preserving meaningful variation in actors, roles, and local context?

\section{From Case-Specific Model to Configurable Framework}

To address this challenge, we develop a configurable role-based Multi-Agent Computer Model (MACM) within the PRO-CLIMATE project\footnote{The work reported here is part of the EU Horizon PRO-CLIMATE project with grant agreement No. 101137967.}. Rather than treating stakeholders as fixed agent types, the model represents them through combinable roles corresponding to recurring deliberative functions. This supports a clearer mapping from empirical stakeholder analysis to computational behaviour. The design draws on empirical agent specification in socio-ecological modelling, participatory decision modelling \cite{ref_lepira}, opinion dynamics \cite{ref_deffuant}, and role-oriented multi-agent systems \cite{ref_ferber}.

The model was upscaled from a case-specific implementation into a general framework through two linked changes. First, it moved from a more granular individual-level abstraction to a stakeholder-group representation. This reduces micro-level detail, but makes model configuration feasible across Living Labs where comparable individual-level data are not available. Second, the model externalises context-sensitive inputs such as stakeholder sets, role assignments, network structure, proposal criteria, expert coverage, and contextual parameters. The behavioural logic remains fixed, while empirical configuration is supplied through structured input files\footnote{These inputs can be derived from stakeholder mapping, workshops, interviews, document analysis, and elicited network ties.}.

A shared computational core supports comparison across Living Labs, while externalised inputs preserve sensitivity to local governance arrangements, stakeholder constellations, and proposal characteristics. Fixed mechanisms include the four-phase process, aggregation of conflicting expert evaluations, stance updating during influence rounds, and the final decision logic. Configurable elements include stakeholders, role assignments, criteria, network structure, and exogenous parameters such as urgency and public visibility. The result is a reusable experimental framework bounded by empirical inputs rather than a one-off case study or a toy model.

\section{Role-Based Design and Simulation Flow}

Each real-world stakeholder is represented as one computational agent that may play one or more roles. Roles define phase-specific behaviour rather than social identity. This matters because stakeholders in real governance settings often have overlapping responsibilities. 
A municipal department, for example, may evaluate institutional feasibility, disseminate information, and also participate in formal decision-making. Fixed agent types would either split such a stakeholder into artificial entities or collapse distinct functions into a single behaviour. A role-based representation avoids both problems and keeps the mapping from stakeholder analysis to model design transparent.

\begin{figure}[t]
\centering
\includegraphics[width=0.99\textwidth]{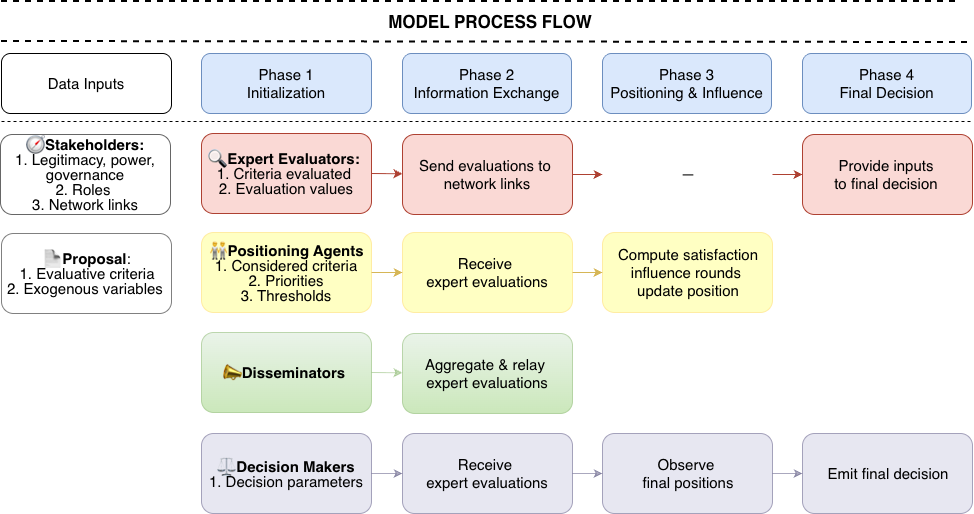}
\caption{Overview of the configurable MACM: Living-Lab inputs define the context, while shared model logic runs four phases with role-specific activation.
}
\label{fig:flow}
\end{figure}

The model uses four principal roles. \textbf{Expert Evaluators} assess criteria such as environmental impact, policy fit, feasibility, cost, or benefit. \textbf{Disseminators} relay or aggregate evaluations through the network. \textbf{Positioning Agents} interpret available information through weighted priorities and form support or opposition. \textbf{Decision-Makers} issue the final collective outcome. Because roles are combinable, the same stakeholder can participate in several stages without duplicating entities in the model.
Figure~\ref{fig:flow} shows the overall simulation flow. In \emph{initialization}, the model loads stakeholder attributes, role assignments, proposal criteria, thresholds, and network ties. In \emph{information exchange}, expert evaluations travel through the network and conflicting values are aggregated through legitimacy-weighted rules. In \emph{positioning and influence}, positioning agents compute satisfaction from available evaluations and weighted criteria, then revise their stance through ego-network influence. In \emph{final decision-making}, decision-makers combine official expert input and observed stakeholder positions to compute utilities for accepting, rejecting, or revising the proposal.

A further design distinction separates \emph{evaluative criteria} from \emph{exogenous variables}. Evaluative criteria---such as environmental, social, economic, vulnerability, policy-fit, and institutional dimensions---enter assessment directly. Exogenous variables such as urgency or public visibility do not directly change satisfaction, but shape the process by affecting engagement and the number of influence rounds \cite{ref_lepira}. Stakeholder priorities are elicited through pairwise comparisons among criteria and translated into approximate weights, which preserves uncertainty from workshop-based elicitation while enabling formal simulation.

\section{Implications for Social Simulation}

The model offers a role-based architecture that is interpretable, empirically configurable, and reusable across cases. It improves institutional realism by representing differentiated stakeholder functions, improves generalisability by keeping mechanisms stable across Living Labs, and supports process tracing through intermediate outputs for information reception, stance formation, influence, and final decision.

A central strength of the approach is that it separates stable deliberative functions from case-specific empirical content. In other words, the model does not need to be redesigned for each Living Lab; instead, roles, parameters, and contextual inputs can be reconfigured while preserving the same core interaction structure. This is especially useful in comparative social simulation, where one often wants to examine how differences across settings emerge from stakeholder constellations, institutional conditions, or communication patterns rather than from hidden changes in the model itself.
The role-based formulation also strengthens explanation. Because the model records who evaluated what, who transmitted which information, how positions changed, and how the final choice was assembled, it supports process-oriented analysis rather than only reporting end states. This matters for climate adaptation governance, where contested outcomes often depend not only on preferences, but also on the sequencing and legitimacy of information, the visibility of proposals, the credibility of actors, and the extent to which decision-makers are exposed to social influence. The framework therefore helps connect observable deliberative dynamics to simulated outcomes in a way that is analytically transparent.

This paper reports a design contribution rather than a completed validation study. Future work will instantiate the model in different Living Labs, validate stakeholder roles and parameters with participants, and compare simulated information flows, stance changes, and final decisions with empirical observations.
%In that sense, the contribution is not a single case-specific model, but a transferable modelling design for representing deliberative climate adaptation processes across heterogeneous urban contexts.

\end{document}